# Spin-to-orbital momentum conversion via electrooptic Pockels effect in crystals

Ihor Skab, Yurij Vasylkiv, Ihor Smaga and Rostyslav Vlokh*

Institute of Physical Optics, 23 Dragomanov St., 79005 Lviv, Ukraine

* Corresponding author: vlokh@ifo.lviv.ua

## Abstract

In the present work we have demonstrated a possibility for real-time operation by orbital angular momentum (OAM) of optical beams via Pockels effect in solid crystalline materials. Basing on the analysis of optical Fresnel ellipsoid perturbed by conically shaped electric field, we have shown that the point groups of crystals convenient for the conversion of spin angular momentum (SAM) to OAM should contain a three-fold symmetry axis or a six-fold inversion axis. The results of our experimental studies and theoretical simulations of the SAM-to-OAM conversion efficiency carried out for $LiNbO_3$ crystals agree well with each other.

## 1. INTRODUCTION

The interest of researchers in real-time operation by orbital angular momentum (OAM) of optical beams is caused by novel possibilities appearing in relation with quantum computing, cryptography, and quantum teleportation [1–3]. Following from the principles of singular optics [4], these possibilities appear if one uses spin angular momentums (SAMs) and OAMs of optical beams, which represent quantum quantities. The SAM associated with orthogonal circular polarizations of light can acquire only two values expressed in the units of $\hbar$ as $s = +1$ and $s = -1$ [5], while the OAM value can, in principle, be infinite ($l = 0, \pm 1, \pm 2, ...$)

[6]. Hence, the OAM has some advantages in the information processing, when compare with the SAM, since a single photon has only two distinct spin states and infinitely many distinct OAM states. In such a case, the information can be encoded by multiplying a number of distinguishable states, because a photon can carry arbitrarily large amount of information distributed over its spin and orbital quantum states [7]. Thus, novel possibilities for applications of these quantum properties of photons arise, e.g. utilization of qubits and qudits in the information processing performed in quantum computers. They are capable of considerable increase in the information content which can be simultaneously processed. A problem which should still be solved when realizing quantum photonic encoding consists in a need for developing efficient methods for real-time SAM-to-OAM conversion.

It has been shown [8–11] that so-called q-plates which represent specific liquid-crystalline plates revealing a structural defect in the geometrical center of the plate, with a topological charge equal to unity, facilitate efficient SAM-to-OAM conversion performed with the aid of temperature tuning or electrooptic operation of orientation of the director. In case of propagation of nearly plane circularly polarized light wave through such a plate, the outgoing helical mode acquires the OAM equal to ±2. Notice that liquid crystals have a number of disadvantages, when compared with their solid analogues. For instance, they manifest low response speeds due to their viscosity, are relatively unstable, cannot be used for operation of powerful laser radiation, and often reveal unnecessary nonlinear responses.

Recently we have suggested a method for the SAM-to-OAM conversion that employs solid crystalline materials subjected to torsion stresses [12, 13]. Then the emergent light beam should have the OAM equal to ±1, whereas the topological defect strength associated with optical indicatrix orientation is equal to ±1/2, in terms commonly used for the liquid crystal-based singular optics. As a consequence, the OAM would be described by a set of even quantum numbers $l = 0, \pm 2, \pm 4...$ in case of the q-plates with the unit topological defects [8, 11], and a set of discrete values $l = 0, \pm 1, \pm 2,...$ in case of the solid crystals. In other words, utilization of the latter should increase twice the number of states in which information can be encoded. On the other hand, piezooptic effect associated with torsion stresses is rather difficult to accomplish in practice when designing relevant devices and, moreover, some additional piezoelectric transducers are necessary in order to convert electrical signals to mechanical stresses.

Hence, development of direct, electrically driven, operation of the OAM on the basis of solid crystalline materials represents an important problem. In our last work [14] it has been shown that the electric field with a special configuration applied to electrooptic crystals can lead to appearance of OAM in the outgoing light beam, provided that the incident circularly polarized beam propagates along the optic axis direction. It has also been found that the topological charge of the outgoing helical mode is equal to unity. Furthermore, we have demonstrated on the canonical examples of electrooptic crystals of $LiNbO_3$ and $LiTaO_3$ that the efficiency of the SAM-to-OAM conversion can be gradually operated by the electric field, using a Pockels effect. Then the following question arises: which of the point symmetry groups describing solid crystals are appropriate for the SAM-to-OAM conversion via Pockels effect? This represents the main goal of the present work. Besides, an experimental verification of the electrooptic control of the SAM-to OAM conversion will also be reported.

## 2. BASIC RELATIONS AND RESULTS OF SIMULATION

As already mentioned, we have earlier demonstrated that a spatial distribution of optical birefringence induced by the torsion stresses reveals a singular point of zero birefringence, which belongs to the torsion axis [15–19]. In general, the coordinate distribution of the torsion-induced birefringence has a conical shape. Due to this distribution, the outgoing wave acquires a helical phase and an OAM. It has been shown in [14] that, while searching electrooptic analogues of the torsion-induced spatial birefringence distributions, one should proceed from the following requirements: a crystal has to be non-centrosymmetric and an electric field should be spatially distributed in a specific manner, with a singular value at the line parallel to the beam axis $Z$, which crosses the geometrical center of its $XY$ cross section (here and below the axes of the coordinate system $XYZ$ are parallel to the eigenvectors of the Fresnel ellipsoid). The conditions mentioned are satisfied when a 'conical' spatial distribution of the electric field (see Fig. 1) is created in crystals [14]. Such a distribution can be produced by two circular electrodes attached to the front and back $XY$ faces of a sample. When the electrodes essentially differ by their radiuses (e.g., the radius of one of them tends to zero), the projections $E_1 = E_x$ and $E_2 = E_y$ of the electric field would appear. Then the electric field components are given by the relations

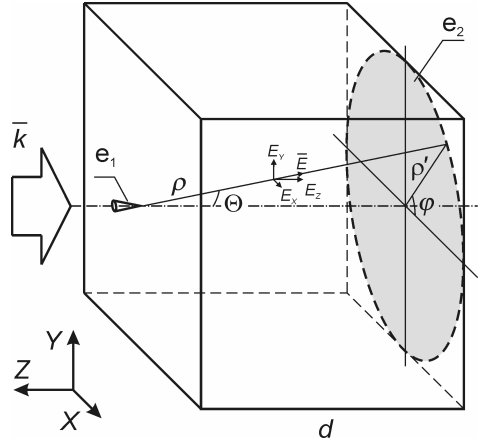

FIG. 1. Schematic presentation of a crystalline plate with circular electrodes $e_1$ and $e_2$, and a conical spatial distribution of electric field produced by these electrodes.

$$E_1 = kX \ , \ E_2 = kY \ , \ E_3 = kZ \ , \tag{1}$$

where

$$k = \frac{U}{d} \frac{Z}{X^2 + Y^2 + Z^2} \ . \tag{2}$$

In the spherical coordinate system defined by $X = r \sin \Theta \cos \boldsymbol{j}$, $Y = r \sin \Theta \sin \boldsymbol{j}$ and $Z = r \cos \Theta$, we obtain

$$E_1 = \frac{U}{d} \frac{\tan \Theta}{1 + \tan^2 \Theta} \cos \boldsymbol{j} \ , \tag{3}$$

$$E_2 = \frac{U}{d} \frac{\tan \Theta}{1 + \tan^2 \Theta} \sin \boldsymbol{j} \ , \tag{4}$$

$$E_3 = \frac{U}{d(1 + \tan^2 \Theta)} \ , \tag{5}$$

where $U$ denotes the applied electric voltage, $d$ the thickness of a crystalline plate, and $E_0 = U / d$. One can see that the $E_1$ and $E_2$ components are equal to zero when $\Theta = 0$ (a case of homogeneous field, with the field lines parallel to the $Z$ axis) and increase with increasing $\Theta$ and decreasing $d$. Notice that the electric field appearing behind the cone limited by the filed lines presented in Fig. 1 is neglected.

The electrooptic Pockels effect is described by the relation $\Delta B_i = r_{ij} E_j$, with $r_{ij}$ being the electrooptic tensor components and $\Delta B_i$ the increments of the optical impermeability tensor. The electrooptic tensor has the following form:

$$
\begin{array}{c|ccc}
 & E_1 & E_2 & E_3 \\
\hline
\Delta B_1 & r_{11} & r_{12} & r_{13} \\
\Delta B_2 & r_{21} & r_{22} & r_{23} \\
\Delta B_3 & r_{31} & r_{32} & r_{33} \\
\Delta B_4 & r_{41} & r_{42} & r_{43} \\
\Delta B_5 & r_{51} & r_{52} & r_{53} \\
\Delta B_6 & r_{61} & r_{62} & r_{63}
\end{array}
, \tag{6}
$$

where the tensor components displayed by grey color are of no practical importance in the present analysis. Since the optical beam propagates along the $Z$ axis, we may consider only the $XY$ cross section of the Fresnel ellipsoid in what follows. The equation of the $XY$ cross section of the optical indicatrix perturbed by the mentioned field configuration reads as

$$
\begin{aligned}
(B_1 + r_{11}E_1 + r_{12}E_2 + r_{13}E_3)X^2 + (B_1 + r_{21}E_1 + r_{22}E_2 + r_{23}E_3)Y^2 \\
+ 2(r_{61}E_1 + r_{62}E_2 + r_{63}E_3)XY = 1
\end{aligned}
, \tag{7}
$$

where the components should $E_1$ and $E_2$ play a major part. Basing on the above analysis, one has to select the point symmetry groups for which the coefficients $r_{11}$, $r_{12}$, $r_{21}$, $r_{22}$, $r_{61}$ and $r_{62}$ do not equal to zero. Moreover, the crystals which belong to these symmetry groups should be optically isotropic or uniaxial in order to fulfill the condition for existence of the singular birefringence point, i.e. a zero birefringence value in the center of the $XY$ cross section. The trigonal and hexagonal groups 3m, 32, 3, $\overline{6}$, and $\overline{6}$m2 belong to such systems.

The common feature of these groups is a three-fold axis (the groups containing the three-fold axis are subgroups of those with the inversion six-fold axis). The non-centro-symmetric cubic groups 23 and $\overline{4}$3m also contain a three-fold axis along [111] directions, although their coefficients $r_{11}$, $r_{12}$, $r_{21}$, $r_{22}$, $r_{61}$ and $r_{62}$ written in the eigen (crystallographic) coordinate system are equal to zero. Nonetheless, if we rewrite the tensor in a coordinate system where the $Z'$ axis becomes parallel, e.g., to [111] direction, these coefficients should not remain zero (notice that the analogical procedure has been used when analyzing the OAM induced via piezooptic effect under crystal torsion [13]). Thus, we start our analysis with the cubic crystals for the case when a nearly plane optical wave propagates along the $Z'$ axis belonging to (110) plane (see Fig. 2), in particular along the three-fold symmetry axis (i.e., the direction [111] of the principal coordinate system).

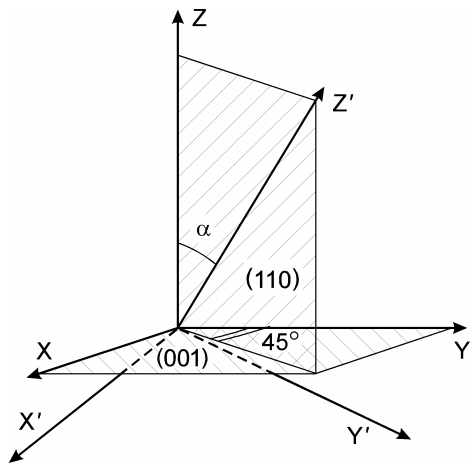

FIG. 2. Schematic representation of rotation of our working coordinate system with respect to the crystallographic one, which is used in order to describe the electric field-induced bire-fringence appearing for the propagation directions defined by angle $\boldsymbol{a}$ .

### A. Cubic crystals of the symmetry groups 23 and $\overline{4}3m$

The electrooptic tensor written in the principal coordinate system for the crystals belonging to the point symmetry groups 23 and $\overline{4}3m$ is given by a single independent electrooptic co-efficient $r_{41}$ ( $r_{41} = r_{52} = r_{63}$ ). However, if one rewrites the tensor in the coordinate system of which the $Z'$ axis is parallel to the direction that belongs to the (110) plane (see Fig. 2, where the angle between the $Z'$ and $Z$ axes is equal to $\boldsymbol{a}$ ), the coefficients under interest are given by

$$r'_{11} = r'_{22} = -\frac{3\sqrt{2}}{4} r_{41} \sin^3 \boldsymbol{a},$$

$$r'_{12} = r'_{21} = r'_{61} = r'_{62} = \frac{\sqrt{2}}{4} r_{41} (1 + 3\cos^2 \boldsymbol{a}) \sin \boldsymbol{a}.$$

(8)

Besides, as seen from Eq. (7), the coefficients $r_{13}$, $r_{23}$ and $r_{63}$ can also induce the birefringence whenever $r_{13} \neq r_{23}$ and $r_{63} \neq 0$. The relations for these tensor components rewritten in the primed coordinate system have the following form:

$$r'_{13} = r'_{23} = -\frac{3}{2} r_{41} \sin^2 \boldsymbol{a} \cos \boldsymbol{a},$$

$$r'_{63} = \frac{1}{2} r_{41} (3\cos^2 \boldsymbol{a} - 1) \cos \boldsymbol{a}.$$

(9)

It is easily seen that $r'_{13}$ is equal to $r'_{23}$, irrespective of the $\boldsymbol{a}$ values. However, we have $r_{63} = 0$ only when $\boldsymbol{a} = 90 \deg$ (the [110] direction) or $54.74 \deg$ (the [111] direction). The dependences of the appropriate electrooptic tensor coefficients on the angle $\boldsymbol{a}$ are displayed in Fig. 3.

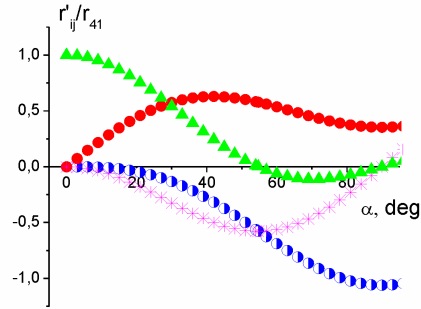

FIG. 3. Dependences of relative electrooptic tensor coefficients $r'_{ij} / r_{41}$ on the angle $\boldsymbol{a}$ : $r'_{11} / r_{41}$ (semi-open circles), $r'_{12} / r_{41}$ (full circles), $r'_{13} / r_{41}$ (crosses), and $r'_{63} / r_{41}$ (full triangles).

The relations for the birefringence and the angle of optical indicatrix rotation may be expressed as

$$\Delta n_{XY'} = -\frac{1}{2}n^3\sqrt{(r'_{11}-r'_{12})^2(E_{X'}-E_{Y'})^2 + 4\left(r'_{12}(E_{X'}+E_{Y'})+r'_{63}E_{Z'}\right)^2}$$

$$= -\frac{1}{2}n^3 E_0 \frac{Z'}{X'^2+Y'^2+Z'^2}\sqrt{(r'_{11}-r'_{12})^2(X'-Y')^2 + 4\left(r'_{12}(X'+Y')+r'_{63}Z'\right)^2}, \tag{10}$$

$$\tan 2z_{Z'} = \frac{2\left(r'_{12}(E_{X'}+E_{Y'})+r'_{63}E_{Z'}\right)}{(r'_{11}-r'_{12})(E_{X'}-E_{Y'})} = \frac{2\left(r'_{12}(X'+Y')+r'_{63}Z'\right)}{(r'_{11}-r'_{12})(X'-Y')}. \tag{11}$$

In general, as seen from Fig. 3, a number of special values of the $\boldsymbol{a}$ angle should be considered separately, namely: $\boldsymbol{a}=0$ deg ( $r'_{11}=r'_{22}=r'_{12}=r'_{21}=r'_{13}=r'_{23}=0$, $r'_{63}=r_{41}$), $\boldsymbol{a}=90$ deg ( $r'_{63}=r'_{13}=r'_{23}=0$, $r'_{12}=r'_{21}=0.35\,r_{41}$, $r'_{11}=r'_{22}=-1.06\,r_{41}$), and $\boldsymbol{a}=54.74$ deg ( $r'_{12}=r'_{21}=-r'_{13}=-r'_{23}=-r'_{11}=-r'_{22}=0.57\,r_{41}$, $r'_{63}=0$ ).

The spatial distributions of the effective birefringence (i.e., the birefringence appearing in an inhomogeneous sample for certain directions of light propagation) and the angle of optical indicatrix rotation simulated for $Bi_{12}TiO_{20}$ crystals taken as an example of cubic electrooptic crystals are presented in Fig. 4 and Fig. 5, respectively. Here the point symmetry group is 23, $n=2.56$, and $r_{41}=5.7\times10^{-12}$ m/V at $\boldsymbol{l}=632.8$ nm [20]. The rest of the parameters are taken to be $R=0.01$ m (the larger electrode radius), $d=0.005$ m (the sample thickness), and $U=5000$ V. Continuous changes in the effective birefringence and the angle of optical indicatrix rotation occurring with changing $\boldsymbol{a}$ angle are illustrated using multimedia. Notice that the spatial distributions of the effective birefringence presented in Fig. 4 have been obtained using integration of the phase difference along the direction of light propagation basing on the function given by formula (10). In other words, the mean value of the phase difference function over the optical path has been derived on the basis of the evident relationship $\Delta\Gamma = -\frac{2\boldsymbol{p}}{\boldsymbol{l}}\int\limits_{\frac{r'd}{R}}^{d}\Delta n_{XY}dZ$. As shown in [14, 19], the mean value of the phase difference function is equal to the effective phase difference calculated using a straightforward Jones matrix approach.

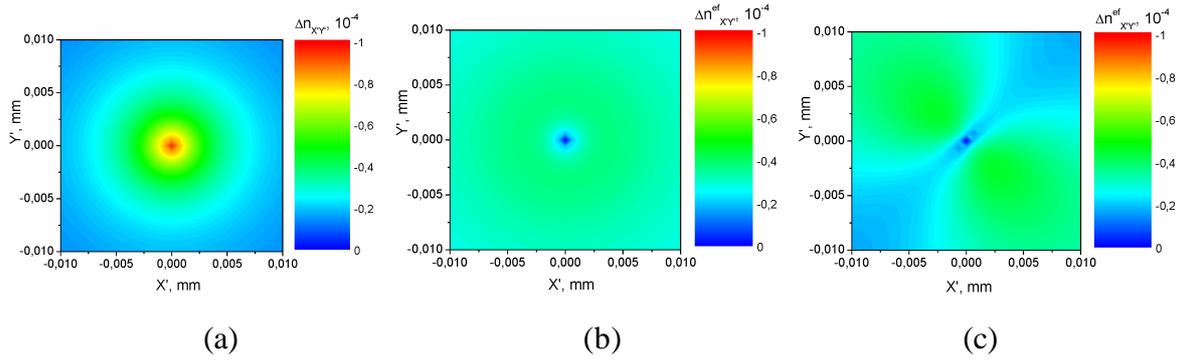

FIG. 4. Spatial distributions of the effective birefringence induced by conically shaped electric field in the $X'Y'$ plane of $Bi_{12}TiO_{20}$ crystals (a – $\alpha = 0 \deg$, b – $\alpha = 54.74 \deg$, c – $\alpha = 90 \deg$) (see also multimedia file 1).

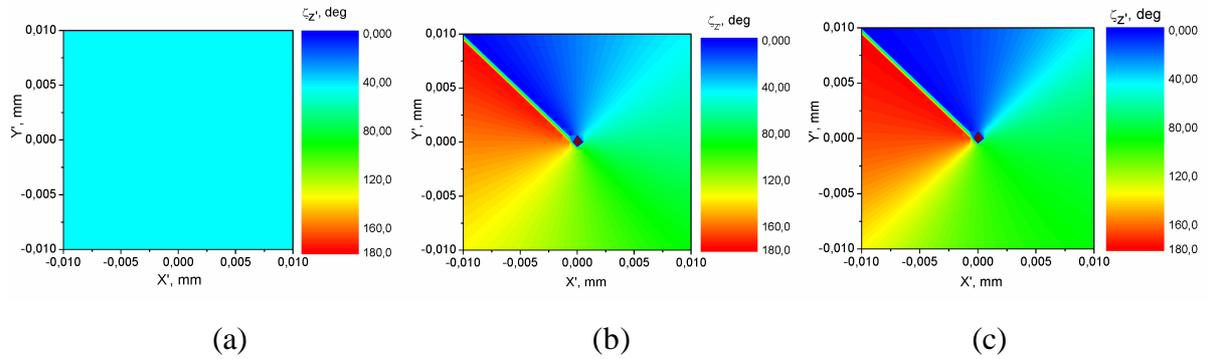

FIG. 5. Spatial distribution of the angle of optical indicatrix rotation induced by conically shaped electric field in the $X'Y'$ plane of $Bi_{12}TiO_{20}$ crystals (a – $\alpha = 0 \deg$, b – $\alpha = 54.74 \deg$, c – $\alpha = 90 \deg$) (see also multimedia file 2).

As seen from Fig. 4(a), at $\alpha = 0 \deg$ the module of the effective birefringence has a maximum in the center of the $X'Y'$ cross section and follows to zero with increasing distance $r'$ (here the polar coordinate system is introduced, with $X' = r'\cos\varphi$ and $Y' = r'\sin\varphi$). The angle of optical indicatrix rotation is equal to $45 \deg$ throughout all the $X'Y'$ cross section (see Fig. 5(a)). In fact, following from Eqs. (8)–(11) we have

$$\Delta n_{XY'} = -n^3 E_0 \frac{Z'^2}{r'^2 + Z'^2} r_{41} = -n^3 E_0 \frac{Z'^2}{r'^2} r_{41} \text{ and } \zeta_{Z'} = 45°, \qquad (12)$$

so that, at $r' = 0$, the module of the induced birefringence reveals a maximum ($\Delta n_{X'Y'}^{ef} = -n^3 E_0 r_{41}$). Hence, the spatial distribution of the effective birefringence has a singular point in the center of the $X'Y'$ cross section. However, this singular point cannot lead to appearance of screw dislocation of the phase front of the emergent beam (as well as appearance of an optical vortex and an OAM) because of lack of any dependence of $z_{Z'}$ (and so of the phase of light wave) on the tracing angle $j$. Moreover, the intensity of light emergent from the system consisting of a sample placed between circular polarizers is not equal to zero at $r' = 0$, being dependent on the electrically induced phase difference.

With increasing $a$ angle (see Fig. 4, Fig. 5, and the multimedia files 1 and 2), the maximum of the effective birefringence is gradually replaced by its minimum ($\Delta n_{X'Y'}^{ef} = 0$). Notice that the coordinates of the minimal effective birefringence correspond to those for which the angle of optical indicatrix rotation has indefinite value. At $a = 54.74 \deg$ (see Fig. 4(b) and Fig. 5(b)), the point that corresponds to zeroes of the effective birefringence and indefinite indicatrix rotation angle occupies the center of the $X'Y'$ cross section. Following from Eqs. (8)–(11), one can write for this case

$$\Delta n_{XY'} = -\sqrt{\frac{2}{3}} n^3 r_{41} E_0 \frac{r'}{r^2} Z', \tag{13}$$

$$\tan 2z_{Z'} = -\frac{(X'+Y')}{(X'-Y')} = -\frac{\cos j + \sin j}{\cos j - \sin j} = \tan\left(\frac{3p}{4} - j\right), \text{ or } z_{Z'} = \frac{3p}{8} - \frac{j}{2}. \tag{14}$$

Thus, the optical indicatrix rotation angle changes twice as slower than the tracing angle $j$, beginning from the initial value $j_0 = 3p/8$ that corresponds to appearance of a pure screw dislocation of the phase front of the emergent light and the OAM equal to unity (see our recent works [12, 13]).

When considering the central point of the $X'Y'$ cross section as a defect on spatial map of the optical indicatrix orientation and using definitions usual for the liquid-crystal terminology, one can see that the topological strength of this defect is equal to $q = \pm 1/2$ and the angle $z_{Z'}$ amounts to half the angle $j$. It is obvious that the sign of the optical indicatrix rotation would reverse whenever the sign of the applied electric field does so. This corre-sponds to a sign reversal of the induced effective birefringence and the optical indicatrix rotation by 90 deg. Contrary to the torsion-induced birefringence [12, 13], the birefringence caused by the electric field reveals nonlinear dependences on the both coordinates $X'$ and $Y'$, which in general are defined by the ratio $r'Z'/r^2$.

The induced effective birefringence reveals a circular distribution in the $X'Y'$ plane (see Fig. 4(b)), being equal to zero at the geometrical center of the $X'Y'$ cross section. A typical spatial distribution of the effective birefringence along the $X'$ axis is shown in Fig. 6. It is evident from Fig. 6 that the module of the effective birefringence increases with increasing $|X'|$ coordinate, tending from a zero value at $X' = 0$ to its maximum at $X'_{max} = \pm 4.0$ mm. Note that the $X'_{max}$ value is independent of either the electrooptic coefficient or the electric voltage applied. Further increase in $X'$ produces decrease in the induced value $\left| \Delta n^{ef}_{XY'} \right|$. The coordinate dependence of the effective birefringence mentioned above is caused by the two mechanisms: (i) increase in $\left| \Delta n^{ef}_{XY'} \right|$ occurring with increasing $|X'|$, due to increasing $E_1$ and $E_2$ projections, and (ii) decrease in $\left| \Delta n^{ef}_{XY'} \right|$ occurring with increasing $|X|$, due to decreasing effective optical path (i.e., the optical path in that part of crystal which is subjected to the electric field).

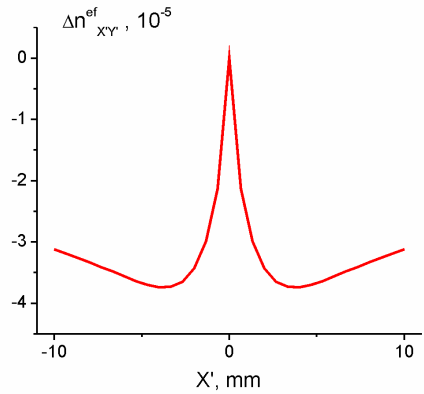

FIG. 6. Dependence of the effective birefringence of $Bi_{12}TiO_{20}$ crystals induced by the electric voltage 5 kV on the $X'$ coordinate.

It has been shown in [14] that the appropriate analytical relationship describing the effective birefringence distribution correspond to the relation for the mean value of the function

$$\Delta n_{XY'}^{ef} = -\frac{n^3 r_{41} E_0 R \boldsymbol{r}'}{\sqrt{6} d (R - \boldsymbol{r}')} \ln \frac{(\boldsymbol{r}'^2 + d^2) R^2}{(R^2 + d^2) \boldsymbol{r}'^2} \ . \tag{15}$$

As a consequence, the minimization procedure yields the following radius that corresponds to the extreme $\Delta n_{XY'}^{ef}$ value:

$$\boldsymbol{r}'_{\max} = d \frac{\sqrt{-W\left(-\dfrac{2R^2 e^{-2}}{R^2 + d^2}\right)\left(W\left(-\dfrac{2R^2 e^{-2}}{R^2 + d^2}\right) + 2\right)}}{W\left(-\dfrac{2R^2 e^{-2}}{R^2 + d^2}\right) + 2} , \tag{16}$$

where $W\left(-\dfrac{2R^2 e^{-2}}{R^2 + d^2}\right)$ implies the *LambertW* function. This is why the mentioned radius depends only on the geometrical parameters of the sample and the electrodes.

With further increase in the angle $\boldsymbol{a}$ (see Fig. 4 and Fig. 5), the ring of the maximum effective birefringence values with the unchangeable height is gradually transformed into the ring with changeable height, the latter being dependent on the tracing angle $\boldsymbol{j}$ (see Fig. 7). At $\boldsymbol{a} = 90 \deg$, the relations for the birefringence and the optical indicatrix rotation angle are given by the formulas

$$\Delta n_{XY'} = -\frac{1}{2\sqrt{2}} n^3 r_{41} E_0 \frac{\boldsymbol{r}' Z'}{\boldsymbol{r}^2} \sqrt{5 - 6\sin \boldsymbol{j} \cos \boldsymbol{j}} \ , \tag{17}$$

$$\tan 2\boldsymbol{z}_{Z'} = -\frac{X' + Y'}{2(X' - Y')} = -\frac{\cos \boldsymbol{j} + \sin \boldsymbol{j}}{2(\cos \boldsymbol{j} - \sin \boldsymbol{j})} = \frac{1}{2} \tan(3\boldsymbol{p} / 4 - \boldsymbol{j}),$$

$$\boldsymbol{z}_{Z'} = \frac{1}{2} \arctan\left(\frac{1}{2} \tan(3\boldsymbol{p} / 4 - \boldsymbol{j})\right). \tag{18}$$

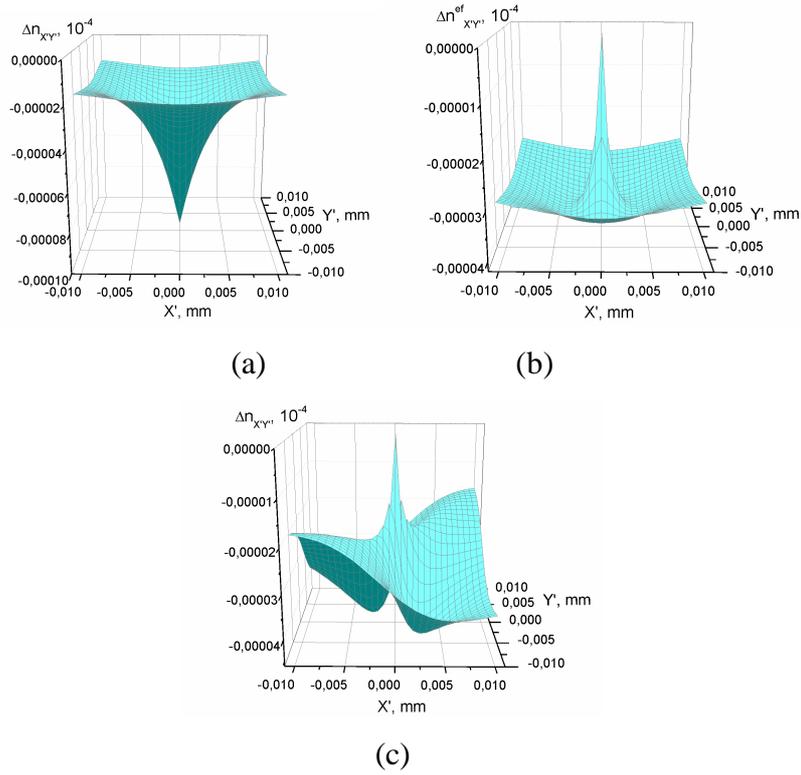

FIG. 7. Spatial distributions of the effective birefringence induced in $Bi_{12}TiO_{20}$ crystals by the electric voltage 5 kV at $\boldsymbol{a} = 0$ deg (a) $\boldsymbol{a} = 54.74\deg$ (b), and $\boldsymbol{a} = 90\deg$ (c).

Eq. (17) is similar to Eq. (13), with the only difference that the extra factor $\sqrt{5 - 6\sin\boldsymbol{j}\,\cos\boldsymbol{j}} = \sqrt{5X'^2 - 6X'Y' + 5Y'^2}$ appearing in Eq. (17) represents the relation for the oval. Hence, the total effective birefringence distribution is given by multiplication of the function of the oval and the function $\Delta n_{X'Y'} \sim E_0 \boldsymbol{r'}Z'/\boldsymbol{r}^2$, which describes a crater-like spatial birefringence distribution (see Fig. 4(b) and Fig. 6). Let us remind that such spatial birefringence distribution produce a mixed screw-edge dislocation of the phase front of the outgoing optical wave (see [13, 18, 21]). This is a reason why the dependence of the optical indicatrix rotation angle on the tracing angle (see Eq. (18) and Fig. 5(c)) is nonmonotonic. In other words, in the case of $\boldsymbol{a} = 90\deg$ we will deal with the mixed screw-edge dislocation of the phase front of light emerging from the system consisting of a sample placed between the crossed circular polarizers. Finally, the changes in the angle $\boldsymbol{a}$ from the zero value to 90 deg should lead to appearance of the screw dislocation of the phase front, which occupies the center of the $X'Y'$ cross section at $\boldsymbol{a} = 54.74\deg$. Further increase in $\boldsymbol{a}$ results in a mixed center of the $X'Y'$ cross section at $a = 54.74\,\mathrm{deg}$. Further increase in $a$ results in a mixed screw-edge dislocation. The pure screw dislocation of the phase front, and so the optical vortex and the OAM, can be observed when the circularly polarized light propagates along the [111] direction in the cubic crystals of the point symmetry groups 23 and $\overline{4}3\mathrm{m}$, and the conically shaped electric field is applied.

*B. Trigonal and hexagonal crystals of the point symmetry groups 3m, 32, 3, $\overline{6}$, and $\overline{6}\mathrm{m}2$*

Now we write out the relations for the birefringence and the rotation angle of optical indicatrix induced in the crystals belonging to the point symmetry groups 3m ($\mathrm{m} \perp X$), 32 ($2 \parallel Y$), and $\overline{6}\mathrm{m}2$ ($\mathrm{m} \perp X$) in case when the conically shaped electric field is applied along the optic axis and the light propagates along the same direction:

$$\Delta n_{XY} = -n_o^3 r_{22} E_0 \frac{Z\sqrt{X^2+Y^2}}{X^2+Y^2+Z^2} = -n_o^3 r_{22} E_0 \frac{\boldsymbol{r}'}{\boldsymbol{r}^2} Z , \qquad (19)$$

$$\tan 2\boldsymbol{z}_Z = \frac{X}{Y} = \cot \boldsymbol{j} \ , \text{ or } \boldsymbol{z}_Z = \frac{\boldsymbol{p}}{4} - \frac{\boldsymbol{j}}{2} , \qquad (20)$$

where $n_o$ means the ordinary refractive index. These relations are almost the same as Eqs. (13) and (14) that describe the birefringence and the optical indicatrix rotation in the cubic crystals, provided that the electric field is applied along the [111] direction and the light propagates in the same direction (see Fig. 4(b) and Fig. 5(b)). Hence, now we will deal with a canonical vortex of which OAM is equal to unity. On the other hand, for the crystals belonging to the point symmetry groups 3 and $\overline{6}$ the corresponding relations have somewhat different form:

$$\Delta n_{XY} = -n_o^3\sqrt{r_{11}^2+r_{22}^2}\,E_0 \frac{Z\sqrt{X^2+Y^2}}{X^2+Y^2+Z^2} = -n_o^3\sqrt{r_{11}^2+r_{22}^2}\,E_0 \frac{\boldsymbol{r}'}{\boldsymbol{r}^2} Z , \qquad (21)$$

$$\tan 2\mathbf{z}_Z = -\frac{r_{22}X + r_{11}Y}{r_{11}X - r_{22}Y} = -\frac{r_{22}\cos\mathbf{j} + r_{11}\sin\mathbf{j}}{r_{11}\cos\mathbf{j} - r_{22}\sin\mathbf{j}}, \text{ or } \mathbf{z}_Z = \frac{\mathbf{p}}{2} - \mathbf{j}_0 - \frac{\mathbf{j}}{2}, \qquad (22)$$

where $\mathbf{j}_0 = \frac{1}{2}\arctan\frac{r_{22}}{r_{11}}$. The latter case differs from the former only by the initial indicatrix rotation angle $\mathbf{j}_0$, which is now defined by the ratio of electrooptic tensor components.

## 3. SAM-to-OAM CONVERSION IN LITHIUM NIOBATE CRYSTALS

Lithium niobate crystals LiNbO$_3$ are canonical electrooptic materials that belong to the symmetry group 3m. Let us analyze the appearance of the OAM in the emergent light beam that has passed through an optical system consisting of a right-handed circular polarizer, a crystalline sample subjected to the electric field of conical configuration, and a left-handed circular polarizer. Notice that the incident beam can be represented by a nearly plane wave and the SAM equal to $S^{inc} = -\hbar$. The electric field of the emergent light may be written as

$$E^{out}(\mathbf{r'}, \mathbf{j}) = E_A \cos\frac{\Delta\Gamma(\mathbf{r'})}{2}\begin{bmatrix} 1 \\ \pm i \end{bmatrix} + iE_A \sin\frac{\Delta\Gamma(\mathbf{r'})}{2}e^{\pm i2qj \pm i2a_0}\begin{bmatrix} 1 \\ \mp i \end{bmatrix}, \qquad (23)$$

where $2q = m = 1$ is the helicity number and $E_A$ the wave amplitude. The first term in Eq. (23) describes the plane wave with the same SAM as in the incident one (i.e., $-\hbar$), while the second one corresponds to the wave with the helical wave front that carries some OAM (see [11]). Since the angular momentum must be conserved, one can write the following relation for the SAM-to-OAM conversion:

$$J^{inc} = J^{out} + M, \qquad (24)$$

where $J^{inc} = S^{inc} = -\hbar$ is the total angular momentum of the incident photon, $J^{out} = L^{out} + S^{out} = -2q\hbar + \hbar = 0$ the total angular momentum of the emergent photon ($S^{out} = +\hbar$, $L^{out} = -2q\hbar$), and $L^{out}$ the OAM of the latter. The mechanical angular momen-tum transferred to the crystalline sample due to the Beth effect [5] is therefore equal to $M = -\hbar$. However, the latter relation which reflects the fact of conservation of the angular momentum has been written under the condition $\Delta\Gamma = p$, though the phase difference can depend on the module $r'$. In this case one should take into account that the plane wave described by the first term of Eq. (23), with the SAM equal to $-\hbar$, also emerges from the sample. Then the efficiency of the SAM-to-OAM conversion is defined by the ratio

$$h = \frac{I_l^{out}}{I_r^{inc}}, \qquad (25)$$

where $I_r^{inc}$ is the intensity of the right-handed incident wave and $I_l^{out}$ that of the left-handed outgoing wave.

The *XY* intensity distribution for the outgoing beam can be calculated using the Jones matrices (see Ref. [12] for more details). The appropriate spatial intensity distributions calculated for different electric voltages are presented in Fig. 8, while the dependences of the efficiency $h$ on the voltage are given in Fig. 9 (here the values $r_{22} = 3.4 \times 10^{-12}$ m/V and $n_o = 2.286$ [22] are taken). As seen from Fig. 8 (a to c), the outgoing light beam corresponds to a pure doughnut mode that bears the OAM. The efficiency of the SAM-to-OAM conversion calculated theoretically reaches ~ 30% at ~ 19 kV for LiNbO$_3$ crystals, if the beam radius equals to $r' = R$. This efficiency can be further increased by decreasing the light beam radius.

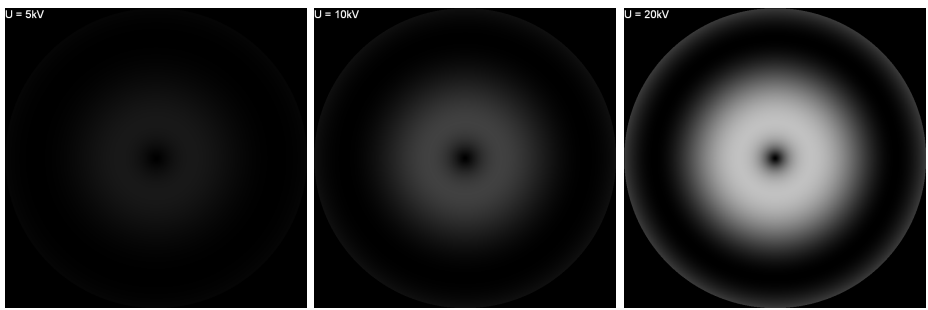

(a)                              (b)                              (c)

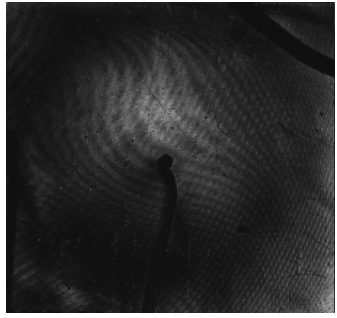

(d)

FIG. 8. Spatial intensity distributions for the beam of 8.5 mm radius emergent from the system consisting of orthogonal circular polarizers and a sample of LiNbO$_3$ in between: a – 5.0, b – 10.0, and c – 20.0 kV (simulation results), d – 10.0 kV (experimental result).

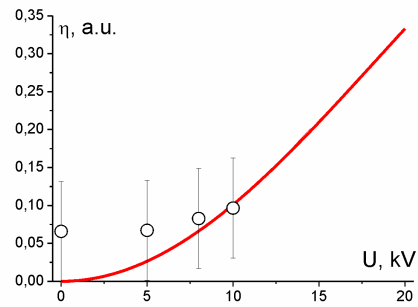

FIG. 9. Dependences of efficiency of the SAM-to-OAM conversion on the electric voltage applied to LiNbO$_3$ crystals: curve corresponds to calculations and points to experimental results.

Lithium niobate crystal sample used in our experiment was prepared as a parallelepiped, with its faces perpendicular to the $Z$ axis. The sample had a thickness of 8.5 mm along the $Z$ axis. The electric field was applied as shown in Fig. 1. The electrode with the smaller radius was fabricated using a metallic wire with the diameter of 0.2 mm, while the electrode of the larger radius ($R = 8.0$ mm) was made as a thin glass plate coated by a transparent conducting SnO layer. The light of a He-Ne laser (the wavelength of $I = 632.8\,\text{nm}$) propagated along the same $Z$ axis which represents the optic axis of our crystal. An experimental setup for the birefringence measurements is shown in Fig. 10. We made the probing beam circularly polarized, since it is not sensitive to the orientation of optical indicatrix. The angle between the principal axes of a quarter-wave plate and the transmission direction of polarizer 8 was equal to 45 deg. In this case the sample can be modeled a linear phase retarder, for which the dependence of the output intensity $I$ on the analyzer azimuth $\boldsymbol{b}$ is expressed as

$$I = \frac{I_0}{2}\left[1 + \sin\Delta\Gamma \sin 2\left(\boldsymbol{b} - \boldsymbol{z}_Z\right)\right] = C_1 + C_2 \sin 2\left(\boldsymbol{b} - C_3\right), \qquad (26)$$

where $\boldsymbol{z}_Z$ is the orientation angle of the optical indicatrix and $\Delta\Gamma = 2\boldsymbol{p}\Delta n_{XY} d / \boldsymbol{l}$ the optical phase difference. After recording and filtering an output image, azimuthal dependences of the intensity $I$ were fitted by the sine function for every pixel of the image, with the fitting coefficients

$$C_1 = \frac{I_0}{2}, \ \ C_2 = \frac{I_0}{2}\sin\Delta\Gamma, \ \ C_3 = \boldsymbol{z}_Z. \qquad (27)$$

It is seen that the optical phase difference $\Delta\Gamma$ is determined by the coefficients $C_1$ and $C_2$:

$$\sin\Delta\Gamma = C_2 / C_1, \qquad (28)$$

while the angular orientation of the intensity minimum is determined by the orientation of principal axis $\zeta_Z$ of the optical indicatrix and the coefficient $C_3$. Hence, fitting of dependences of the light intensity behind the analyzer upon the azimuth for each pixel of the sample image enables constructing 2D maps of optical anisotropy parameters of the sample under study (the optical phase difference and the optical indicatrix orientation).

While studying the optical indicatrix rotation, we exploited an imaging polarimetric setup (Fig. 10), in which linearly polarized incident light was used. An analyzer 10 was placed into the extinction position with respect to a polarizer. The crossed polarizers were simultaneously rotated by 180 deg with the steps of 5 deg. The rotation stage angles corresponding to minimums of the transmitted light intensity detected by a CCD camera were ascribed to the extinction positions (i.e., to the orientation angles of optical indicatrices in different parts of the image).

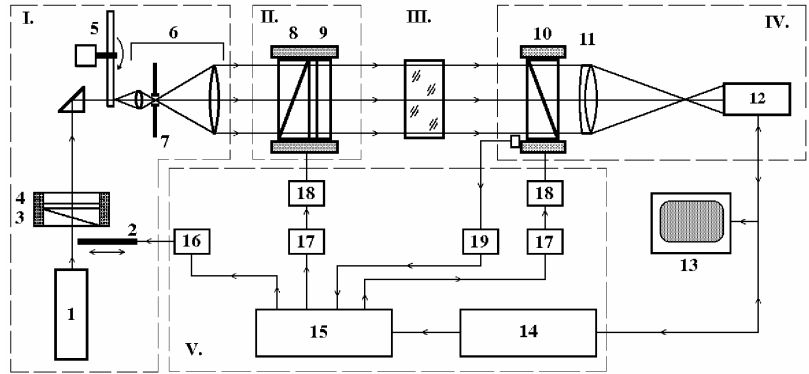

FIG. 10. Scheme of our imaging polarimeter (I – light source section; II – polarization generator; III – sample section; IV – polarization analyzer; and V – controlling unit): 1 – He-Ne laser; 2 – ray shutter; 3, 8 – polarizers; 4, 9 – quarter-wave plates; 5 – coherence scrambler; 6 – beam expander; 7 – spatial filter; 10 – analyzer; 11 – objective lens; 12 – CCD camera; 13 – TV monitor; 14 – frame grabber; 15 – PC; 16 – shutter's controller; 17 – step motors' controllers; 18 – step motors; 19 – reference position controller.

Finally, when checking appearance of the optical vortices, we used a wide circularly polarized incident beam propagating along the $Z$ axis of LiNbO$_3$ subjected to conically shaped electric field. A circular polarizer was used behind a crystalline sample, with its circularity sign opposite to that of the input polarizer. Then an additional quarter-wave plate was placed between a sample section III and an analyzer 10.

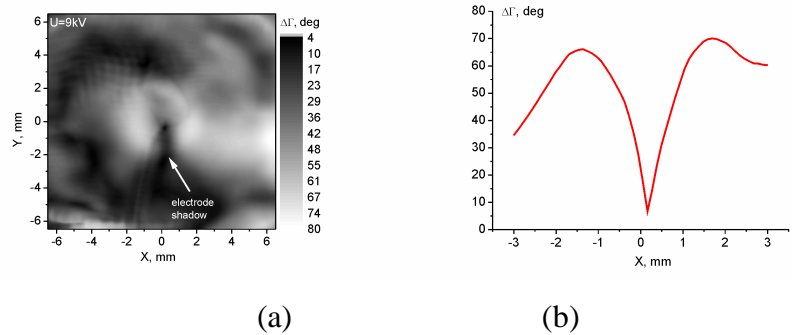

(a)                    (b)

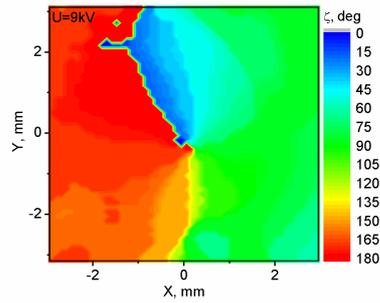

(c)

FIG. 11. Experimental distributions of the phase difference induced by electric voltage $U = 9$ kV in LiNbO$_3$ crystals in the $XY$ plane (a) and along the $X$ axis (at $Y = 0$) (b), and experimental distributions of the optical indicatrix rotation angle in the $XY$ plane.

As seen from Fig 8 (d), the shape of the doughnut mode obtained experimentally agrees well with that estimated theoretically (Fig. 8 (b)). The central dark spot where the optical vortex is located (Fig. 8(b, d)) corresponds to the coordinates of zeroes of the phase difference (Fig. 11(a, b)) and to the singular point of the optical indicatrix rotation angle (Fig. 11(c)). Here the map of the optical indicatrix orientation in the $XY$ plane is characterized by the topological defect with the strength equal to ½ (the optical indicatrix rotation angle is twice as smaller when compared with the tracing angle), so that the vortex transferred by the outgoing beam should be characterized by the OAM equal to unity. The points in Fig. 9 correspond to the data obtained experimentally, whereas the curve gives the efficiency of the SAM-to-OAM conversion calculated theoretically following from the Jones matrix approach and the known values of electrooptic coefficients for the LiNbO$_3$ crystals.

The efficiency of the SAM-to-OAM conversion increases with increasing electric voltage, making it possible to operate the conversion efficiency using the Pockels effect in solid crystalline materials. Somewhat high experimental noise seen in Fig. 9 is mainly caused by the initial optical inhomogeneity of our LiNbO$_3$ crystals. Nonetheless, one can notice that the experimental results obtained by us are in fairly good agreement with the theoretical consideration presented above. The efficiency of the SAM-to-OAM conversion reached in our experiment is about 10% at 10 kV.

## 4. CONCLUSIONS

In the present work we have demonstrated a possibility for real-time operation by the OAM of optical beams based on the Pockels effect in solid crystalline materials. Using the analysis of optical indicatrix perturbed by conically shaped electric field we have proven that the point symmetry groups of crystals convenient for the SAM-to-OAM conversion should contain the three-fold symmetry axis or the six-fold inversion axis. It has also been shown that, in the cubic crystals of the symmetry groups 32 and $\overline{4}3m$, a canonical vortex (i.e., a screw dislocation of the phase front) should appear whenever the light propagates along the [111] direction, the conical electric field is applied along the same direction, and the sample is placed between the crossed circular polarizers. The simultaneous changes in the directions of light propagation and electric field result in displacement of the optical field singularity out of the beam and/or to replacement of the pure screw dislocation by the mixed screw-edge dislocation of the phase front.

For the crystals belonging to the trigonal and hexagonal systems, the application of conically shaped electric field along the optic axis leads to appearance of the pure screw dislocation of the phase front, with the unit topological charge. The results of the experimental studies and the theoretical simulations of the SAM-to-OAM conversion efficiency carried out for the case of $LiNbO_3$ crystals are shown to be in good agreement. The fact of appearance of the pure doughnut mode under the action of electric field in $LiNbO_3$ is experimentally detected. Moreover, we have checked experimentally that the efficiency of the SAM-to-OAM conversion can reach about 10% at the electric voltage of 10 kV. The results presented in this work open a possibility for direct operation by the OAM of optical beams using electric fields via the Pockels effect in the solid crystalline materials.